\documentclass[a4paper,11pt]{article}
\usepackage{pos}

\usepackage{amssymb}

\newcommand{\ev}[1]{\left\langle #1 \right\rangle}
\newcommand{\Var}{\mathrm{Var}}
\newcommand{\Cov}{\mathrm{Cov}}
\newcommand{\Op}{\mathcal{O}}
\newcommand{\dOz}{\Delta\Op_0}
\newcommand{\Stein}{\mathcal{T}_r}
\newcommand{\Lang}{\mathcal{L}_r}
\newcommand{\KL}{\mathrm{KL}}
\newcommand{\dt}{\delta\tau}

\title{A variational framework for variance reduction in lattice field theory}
\ShortTitle{A variational framework for variance reduction}

\author*[a]{Pietro Butti}
\author[b]{Guilherme Catumba}
\author[c]{Alessandro Nada}
\author[a,c]{Louis Spatscheck}
\affiliation[a]{Quantum Theory Centre ($\hbar$QTC) at D-IAS and IMADA, University of Southern Denmark, Campusvej 55, 5000 Odense, Denmark}
\affiliation[b]{Dipartimento di Fisica Teorica "Giuseppe Occhialini", 
Università degli Studi di Milano-Bicocca 
Piazza dell'Ateneo Nuovo, 1 - 20126, Milano
}
\affiliation[c]{Dipartimento di Fisica Teorica, 
Università degli Studi di Torino 
Via Pietro Giuria, 1 - Torino
}
\emailAdd{pbutti@qtc.sdu.dk}

\abstract{%
The signal-to-noise problem limits the reach of many lattice calculations. We
present a variational framework that recasts it as a transport problem: the loss
of signal reflects a mismatch between the distribution one samples and the one
needed to measure an observable, and can be reduced by transporting
configurations to close that gap. The optimal transport is typically determined
either through a stochastic estimator based on Langevin dynamics or by
parametrising it as a normalising flow trained with automatic differentiation.
We discuss how the framework brings these methods under a common variational
principle and present results for scalar theories.%
}

\FullConference{%
The 42nd International Symposium on Lattice Field Theory (LATTICE2026)\\
31 July -- 4 August 2026\\
College Park, Maryland, USA%
}

\begin{document}
\maketitle

\section{Introduction}
\noindent Monte Carlo estimates of Euclidean correlators degrade exponentially with the
separation between the operators: if the signal falls as $e^{-mt}$ while the
variance is set by a lighter state, the relative error grows
exponentially~\cite{ref:ParisiLepage}. This signal-to-noise (StN)
degradation limits many lattice calculations: existing standard methods, e.g. smearing, distillation, GEVP and multilevel schemes aim to improve the quality of the signal at earlier Euclidean times, but do not treat the StN degradation per se.


The problem is reformulated in Ref.~\cite{ref:CatumbaRamos} by writing the two-point
function as the response of a one-point function to an infinitesimal
source, and thus to the evaluation of a derivative of a one-point
function. The conventional method to compute two-point functions is
shown to be a reweighting evaluation of the derivative, which adds an
overlap problem to the degradation of the signal.  Importantly, both
the overlap and the degradation problems can by cured by transporting
the samples until the weights become constant.

Recently two approaches have been pursued in literature: the exact construction of
Ref.~\cite{ref:CatumbaRamos}, based on stochastic Automatic Differentiation (AD), and
trained Normalising Flows (NF)~\cite{ref:Abbott}. Both add to the estimator a
quantity of vanishing mean, i.e. a \emph{control variate}~\cite{ref:controlvariates}. 
We show that all the aforementioned approaches solve
the same variational problem. Minimising the Kullback--Leibler (KL) divergence between
the transported ensemble and the source-deformed target gives one
\emph{pointwise} equation for the transport field, the Stein--Poisson equation,
whose operator generates Langevin dynamics. It supplies the exact solution as a
benchmark, and makes clear that the choice between these approaches is one of
methodology rather than of substance. 

\section{Correlators as source reweighting}\label{sec:rw}
\noindent Consider a Euclidean scalar theory with action $S[z]$, Boltzmann density
$r[z]=e^{-S[z]}/Z_0$, and zero-momentum timeslice operator
$\Op_t=\sum_{\vec x} z(t,\vec x)$; averages over $r$ are denoted $\ev{\cdot}$ and
the connected two-point function is $C(t)=\ev{\Op_t\Op_0}-\ev{\Op_t}\ev{\Op_0}$.
Coupling an infinitesimal source to the timeslice at the origin,
$S_\varepsilon=S-\varepsilon\Op_0$ with density $r_\varepsilon\propto
e^{-S_\varepsilon}$, the correlator is generated by differentiation,
\begin{equation}
  C(t)=\frac{\partial}{\partial\varepsilon}\Big|_{\varepsilon=0}
  \ev{\Op_t}_{r_\varepsilon}.
  \label{eq:response}
\end{equation}
Configurations are drawn from $r$, not $r_\varepsilon$, so the right-hand side is
evaluated by importance reweighting as with
\begin{equation}
  w_\varepsilon[z]=\frac{r_\varepsilon[z]}{r[z]}
  =\exp\!\Big(\varepsilon\Op_0-\log\tfrac{Z_\varepsilon}{Z_0}\Big)
  =1+\varepsilon\,\dOz+\Op(\varepsilon^2),
  \qquad \dOz:=\Op_0-\ev{\Op_0},
  \label{eq:weights}
\end{equation}
and inserting this into $\ev{w_\varepsilon\Op_t}$ reproduces
Eq.~\eqref{eq:response} at order $\varepsilon$. In practice, the exact derivative is
taken numerically by promoting $\varepsilon$ to a formal
series and letting forward-mode AD propagate
it~\cite{ref:CatumbaRamos}. The noise mechanism is now
explicit: the variance of $C(t)$ is dominated by the variance of the weights
$\Var[w_\varepsilon]=\varepsilon^2\Var[\Op_0]+\Op(\varepsilon^3)$, which is induced
by the fluctuation of the source operator.

The general solution to such an overlap problem is to find a change of variables $T(z)$, 
whose distribution is ``closer'' to $r_\varepsilon$ than the original $r$. Typically this is done
by minimising a computable divergence between them. The most used choice is the (reverse) 
KL divergence, which for two densities $r,r_\varepsilon$ reads
\begin{equation}
  \KL(r\Vert r_\varepsilon) = \ev{\log\frac{r}{r_\varepsilon}}_{r}
  \quad\Longrightarrow\quad
  \KL(r\Vert r_\varepsilon)=\tfrac{\varepsilon^2}{2}\Var_r[\Op_0]+\Op(\varepsilon^3),
  \label{eq:kl-chi2}
\end{equation}
the second relation being the baseline mismatch before any transport. Note that
the same quantity $\Var[\Op_0]$ measures both the distance between the two
distributions and the noise of the estimator: reducing one reduces the other.


\section{Transport and the variational problem}\label{sec:flow-exp}
\noindent Let's consider the map $z\mapsto T_\varepsilon(z)$, which transforms samples $r$ into samples of a density $q_\varepsilon$ defined by the change of
variables $q_\varepsilon[T_\varepsilon(z)]=r[z]\,|\det J_{T_\varepsilon}|^{-1}$. In the new variables, the importance weights are given by
\begin{equation}
  \tilde w_\varepsilon\equiv \frac{r_\varepsilon[T_\varepsilon(z)]}{q_\varepsilon[T_\varepsilon(z)]} = \exp\Big(S[z]-S_\varepsilon[T_\varepsilon(z)]
  +\log|\det J_{T_\varepsilon}|-\log\tfrac{Z_\varepsilon}{Z_0}\Big).
  \label{eq:improvedweights}
\end{equation}
which differs from their unimproved version $w_\varepsilon$ in Eq.~\eqref{eq:weights}, by the presence of the Jacobian and a different $\Delta S$, as the target action $S_\varepsilon$ is evaluated onto the new samples $T_\varepsilon(z)$.

Everything that follows is an attempt to satisfy one condition: if the map $T_\varepsilon$ can be
chosen so that the exponent in Eq.~\eqref{eq:improvedweights} takes the same value on every configuration, the
weights are constant, the reweighting is exact with added zero variance, and the StN problem is
solved. This is usually referred to as the \emph{perfect-flow} condition: the change $\Delta S$ in
the action under the map must be compensated, configuration by configuration, by
the change in the volume element. 

We take $T_\varepsilon$ to be generated by a vector field $f$, through
$\mathrm{d}z(\varepsilon)/\mathrm{d}\varepsilon=f(z(\varepsilon))$ with
$z(0)=z_0\sim r$, and work throughout with its linearisation, the
\emph{truncated} map $z\mapsto z+\varepsilon f(z)$ of
Refs.~\cite{ref:CatumbaRamos,ref:Spatscheck}.
Minimising $\KL(q_\varepsilon\Vert r_\varepsilon)$ is the differentiable proxy
for the perfect-flow condition, since
$\KL(q_\varepsilon\Vert r_\varepsilon)=\tfrac12\Var[\tilde w_\varepsilon]
+\Op(\ev{(\tilde w_\varepsilon-1)^3})$ in the small-mismatch regime.

The object that organises the whole $\varepsilon$-expansion of $\KL$ is the \emph{Stein operator}~\cite{ref:SVGD},
\begin{equation}
  \Stein f:=\nabla\!\cdot\!f+f\cdot\nabla\log r
  =\operatorname{tr}J_f-f\cdot\nabla S
  \qquad\text{equivalently}\qquad
  \frac{\mathrm{d}S_\varepsilon}{\mathrm{d}\varepsilon}=-\Stein f \,,
  \label{eq:stein}
\end{equation}
it is the generator of the action deformation induced by the flow. Two properties are used throughout: Stein's identity
$\ev{\Stein f}=0$, valid for every smooth $f$ (conservation of probability along the flow), and the adjoint identity
$\ev{h\,\Stein f}=-\ev{f\cdot\nabla h}$, which moves the operator off the field
and onto a test function as a gradient. 
%
The expansion reads\footnote{
    The derivation combines the Lie--Taylor expansion of $\log p$ along
    the flow with the all-orders expansion of $\log\det J$; see Ref.~\cite{ref:paper}. It is completely 
    general, i.e. does not rely on the linear truncation $z\mapsto z+\varepsilon f$,
    and can be easily extended to cases in which $z$ lives on a Lie group.
}
\begin{equation}
  \KL(q_\varepsilon\Vert r_\varepsilon)=\frac{\varepsilon^2}{2}
  \Bigl\langle\big(\Stein f\big)^2
   - 2f\cdot\nabla\Op_0
  +\Var[\Op_0]\Bigr\rangle+\Op(\varepsilon^3)\,.
  \label{eq:klexp}
\end{equation}
whose minimum $f^\star$ defines the transported samples $z^\star=T_\varepsilon(z)\sim q_\varepsilon$ from which the correlator, extracted as the $\varepsilon$ derivative of $\langle\Op_r\rangle_{q_\varepsilon}$, does not suffer from StN degradation anymore.

\paragraph{The variational problem}
At second order in $\varepsilon$, the whole problem therefore collapses onto the minimisation of a single scalar functional,
\begin{equation}
  f^\star=\arg\min_f \mathcal{F}[f],
  \qquad
  \mathcal{F}[f]:=\ev{\big(\Stein f\big)^2-2f\cdot\nabla\Op_0}
  \label{eq:functional}
\end{equation}
From a computational perspective, the Stein-operator is a scalar observable computed through its definition Eq.~\eqref{eq:stein}, involving the trace of the Jacobian and the field $f$ contracted with the HMC force $\nabla S$, while the second term is $f\cdot\nabla\Op_0=\sum_{\vec x}f(0,\vec x)=:f_0$ for a source linear in the field.
Of the two terms in
Eq.~\eqref{eq:stein}, the Jacobian trace determines the computational bottleneck: for a generic field $f(z)$, the exact trace computational cost scales as $\mathcal{O}(V^2)$. Stochastic sources via the Hutchinson method~\cite{ref:Hutchinson} reduce the computational burden to $\mathcal{O}(V)$, while introducing an underlying stochastic noise floor. On the other hand, $\operatorname{tr}J_f$ is the leading term in the expansion of $\log|\det J_\varepsilon|$, available in a closed form by means of normalising flow-based architectures.\footnote{The trace of $J_f$ can be obtained either by ad-hoc modifications of a standard coupling-layer architecture or by combining NF with forward AD techniques, e.g. truncated polynomials. These approaches are currently being explored by the authors.}

An equivalent form of $\mathcal F$ can be obtained by means of the Stein's adjoint equation~\cite{ref:paper, ref:Spatscheck} as
\begin{equation}\label{eq:kl2loss}
  \mathcal F[f] = \ev{\operatorname{tr}\big(J_f^{2}\big)+f^{\top}H_S\,f-2\,f_0}\,,
\end{equation}
where $H_S:=\nabla^2S$ denotes the
Hessian of the lattice action contracted with $f$ via the Hessian-vector product.


$\mathcal{F}$ is quadratic with a positive-semidefinite quadratic form, so the
minimisation is convex and its stationarity condition closes. 
The adjoint Stein identity yields the associated Euler-Lagrange equation~\cite{ref:paper}, which reads
\begin{equation}
  \boxed{\;\Stein f^\star+\dOz=0\;}
  \label{eq:steinpoisson}
\end{equation}
We will refer to it as the \textbf{Stein-Poisson equation}, which holds configuration by configuration.
Comparing with Eq.~\eqref{eq:weights}, its content is transparent:
$\dOz$ is exactly the $\Op(\varepsilon)$ fluctuation of the log-weights, so
Eq.~\eqref{eq:steinpoisson} demands that the action deformation generated by the
transport (i.e. $\Stein f$) cancels the source fluctuation \emph{pointwise}. It is the perturbative
realisation of the perfect-flow condition above: at the optimum the weight
variance is pushed from $\Op(\varepsilon^2)$ to $\Op(\varepsilon^4)$.


The two formulations are equivalent. Using the adjoint identity at $h=\dOz$ to
complete the square in Eq.~\eqref{eq:functional} gives
$\mathcal{F}[f]+\Var[\Op_0]=\ev{(\Stein f+\dOz)^{2}}$, so that minimising the
divergence is the same as minimising the mean squared violation of
Eq.~\eqref{eq:steinpoisson}. Both are therefore extremised by the same field,
unique up to the kernel of $\Stein$, and the equivalence holds exactly to the
extent that the adjoint identity does. 

\paragraph{A control-variate estimator.}
Repeating the reweighting computation of Sec.~\ref{sec:rw}, now including the transport, gives the per-configuration improved estimator
\begin{equation}
  \hat A_t=\dOz\,\Op_t+(\Stein f)\Op_t+f\cdot\nabla\Op_t,
\end{equation}
whose expectation is $C(t)$ for \emph{every} $f$ by the adjoint identity with $h=\Op_t$.
Thus, the transport contributes an exactly zero-mean term, i.e.~a \textit{control variate}. For a source linear in the field, $f\cdot\nabla\Op_t=\sum_{\vec x}f(t,\vec x)=:f_t$,
and introducing the \emph{residual} $\rho:=\Stein f+\dOz$,
we obtain
\begin{equation}
  \boxed{\hat A_t=f_t+\rho\,\Op_t},
  \qquad
  \ev{\hat A_t}=C(t)\ \ \text{for every }f .
  \label{eq:rhoform}
\end{equation}
This is the main observable that we ought to compute once $f^\star$ is given. Importantly, $\hat A_t$ can be evaluated without any automatic differentiation. Provided the adjoint Stein identity, Eq.~\eqref{eq:rhoform} is an exact identity: $\hat A_t$ is exactly unbiased for any $f$, independently of how well the transport performs. The residual measures the failure of $f$ to solve Eq.~\eqref{eq:steinpoisson} and controls the residual \emph{variance}, but never the correctness of the estimator. 
Although $f_t$ is cheaper to compute since it requires no divergence, its expectation is generally biased, $\ev{f_t}=C(t)-\ev{\rho\,\Op_t}$. Thus, the framework reduces the variance through the transport while preserving exact unbiasedness: the residual term controls the remaining stochastic fluctuations, rather than compensating for a bias introduced by the method.


Writing the control variate as
$\mathcal{A}_t:=(\Stein f)\Op_t+f_t$, which has exactly vanishing mean, the
variance of the improved estimator decomposes as
\begin{equation}
  \sigma^2[\hat A_t]=\underbrace{\sigma^2_C}_{\text{StN degradation}}
  +\underbrace{\Var[\mathcal{A}_t]}_{\text{suppressed by }\mathcal{F}[f]}
  +\,2\underbrace{\Cov\big[\dOz\Op_t,\,\mathcal{A}_t\big]}
  _{\textstyle =\,-\ev{\Op_t^2f_0+\dOz\,\Op_tf_t}}\;,
  \label{eq:vardecomp}
\end{equation}
with $f_0=\sum_{\vec x}f(0,\vec x)$, the identity under the last brace following
from the adjoint identity with $h=\dOz\Op_t^2$. The first two terms are non-negative, so \emph{all} of the improvement comes from the
covariance: the transport adds a zero-mean quantity engineered to anticorrelate
with the fluctuations of the standard estimator, and a good $f$ is one for which
that anticorrelation beats the variance the control variate itself carries.
%
%

Two routes follow: solving the elliptic
equation directly through a probabilistic representation of $\Stein^{-1}$, which
needs no parametrisation of $f$ and is the subject of Sec.~\ref{sec:fk}; or
restricting $f$ to a parametric family and minimising either $\mathcal{F}$ in the
form of Eq.~\eqref{eq:kl2loss}, as in Ref.~\cite{ref:Spatscheck} and
Sec.~\ref{sec:ml}, or the squared residual $\rho$ itself.
In Secs.~\ref{sec:fk} and~\ref{sec:ml}, we present preliminary results for both approaches for a standard $\lambda\phi^4$ theory. These results are part of a broader programme that we will present in a forthcoming publication~\cite{ref:paper}.
\section{Route A: the stochastic resolvent}
\label{sec:fk}
In this section, we tackle the Stein-Poisson Eq.~\eqref{eq:steinpoisson}.
The solution is unique only up to $r$-divergence-free fields, along which
$\mathcal{F}$ is exactly flat. Without loss of generality, we remove the degeneracy by restricting to
gradient fields $f=\nabla\varphi$. With such an ansatz, the Stein operator becomes
\begin{equation}
  \Stein\nabla\varphi\equiv\Lang\varphi,
  \qquad
  \Lang:=\nabla^2-\nabla S\cdot\nabla,
  \label{eq:generator}
\end{equation}
i.e.\ $\Lang$ is precisely the generator of overdamped Langevin dynamics $\mathrm{d}Z_\tau=-\nabla S[Z_\tau]\,\mathrm{d}\tau+\sqrt{2}\,\mathrm{d}W_\tau$. This is
the hinge of the construction: the variational problem has become an elliptic
problem whose operator generates the very dynamics that samples the theory. For a
quadratic action the equation closes on a linear ansatz and returns
$f^\star=K^{-1}v$ with $K$ the kinetic operator and $v=\delta_{x_0,0}$: the
optimal transport is the lattice propagator summed over the source timeslice --
configuration-independent, and, projected onto zero momentum, the very correlator
one is trying to measure. This reproduces the transformation of
Ref.~\cite{ref:CatumbaRamos} and provides the benchmark for everything below.

Beyond the free theory, $\Lang$ must be inverted. Because it generates the
Langevin dynamics, its inverse admits a
probabilistic representation, given by the Feynman-Kac formula~\cite{ref:AlbergoKanwar}
\begin{equation}
  \varphi(z)=\int_0^\infty\!\mathrm{d}\tau\;
  \ev{\dOz[Z_\tau]}_\eta\Big|_{Z_0=z},
  \label{eq:fk}
\end{equation}
where $Z_\tau$ solves the Langevin equation in stochastic time $\tau$ started at $Z_0=z$ and the average $\langle\cdot\rangle_\eta$ runs
over an ensemble of \emph{walkers} sharing that initial configuration. The inverse of the
generator is thus an integral over the natural relaxation of the theory. 

\begin{figure}[t]
  \centering
  \includegraphics[scale=0.25]{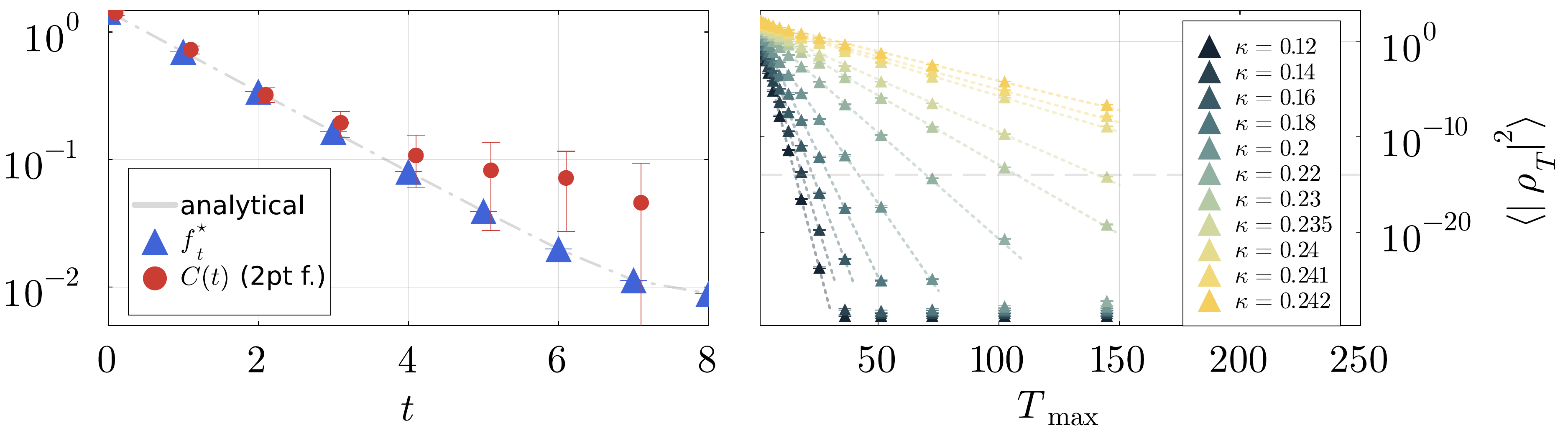}
  \caption{Free theory ($\lambda=0$). \emph{Left:} the transport field
    $f^\star_t$ obtained from Eq.~\eqref{eq:adjointode} follows the analytic
    propagator over the whole range, while the standard estimator of $C(t)$
    loses the signal once its error exceeds it. \emph{Right:} the residual
    $\ev{\rho_T^2}$ against the horizon for a range of hopping parameters; the
    decay is exponential, at a rate that slows as criticality is approached. Dotted lines are exponential fit.
    }
  \label{fig:residual}
\end{figure}

As $f=\nabla \varphi$, we need the gradient of Eq.~\eqref{eq:fk}. Differentiating under the integral and
using that the additive noise drops out, the required object is obtained most
efficiently by the adjoint sensitivity method~\cite{ref:AlbergoKanwar,ref:Li}:
with $\xi=T-\tau$, one solves the deterministic equation
\begin{equation}
  \frac{\mathrm{d}A^\xi}{\mathrm{d}\xi}=-\nabla^2S[Z_{T-\xi}]A^\xi+v,
  \qquad A^0=0,
  \qquad
  f^\star(z)=\lim_{T\to\infty}\ev{A^T}_\eta .
  \label{eq:adjointode}
\end{equation}
The algorithm is a forward pass integrating the Langevin trajectory followed by a
backward pass reading it in reverse; the only ingredient beyond the force is a
Hessian-vector product, and the dominant cost is storing the trajectory. 
In the free theory the outcome can be
checked against the closed form obtained above: Fig.~\ref{fig:residual} (left) shows
$f^\star_t$ tracking the analytic propagator well past the point where the
standard estimator of $C(t)$ has lost its signal.

\begin{figure}[t]
  \centering
  \includegraphics[scale=0.42]{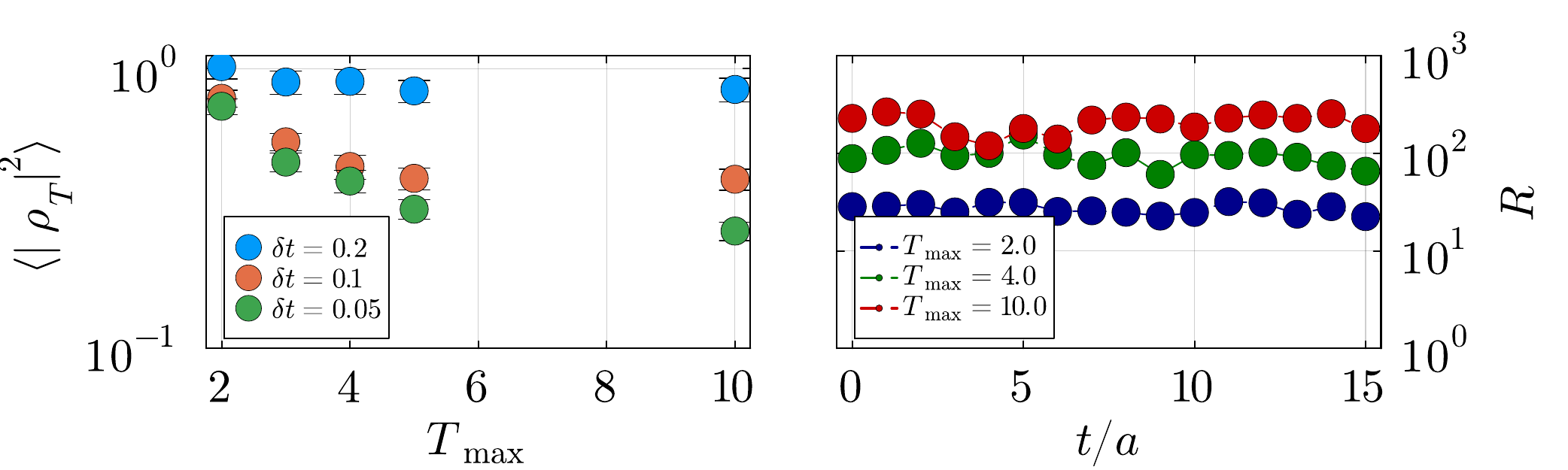}
  \caption{Interacting theory ($\kappa=0.2, \lambda=0.1, V=16\times16$). \emph{Left:} the residual $\ev{\rho_T^2}$ against the horizon at three
    Langevin step sizes; the saturated values are the floors $c(\dt)$ of
    Eq.~\eqref{eq:ledger}. \emph{Right:} the reduction factor
    $R(t)=\sigma^2[U_t]/\sigma^2[\hat A_t]$ of Eq.~\eqref{eq:master} computed at $\delta t=0.05$ against the
    Euclidean separation, at three horizons $T$.}
  \label{fig:money}
\end{figure}

\paragraph{The residual and the reduction factor.}
The representation~\eqref{eq:fk} is exact; in practice it is (i.) truncated at a
finite ``horizon'' $T$, (ii.) integrated at finite step size $\dt$, and (iii.) averaged over
finitely many walkers. As evident in Eq.~\eqref{eq:rhoform}, none of these produces a bias in $\hat A_t$, but all of them leave a non-zero residual
$\rho_T$, and it is worth stating quantitatively what that costs. Since $f_t$
fluctuates only weakly, the variance of the improved estimator is carried by the
$\rho_T\Op_t$ term of Eq.~\eqref{eq:rhoform}, and at large separation the source
leg and the sink decorrelate, so that the four-point averages in
Eq.~\eqref{eq:vardecomp} factorise. The common factor $\ev{\Op_t^2}$ then cancels
in the ratio against the unimproved estimator $U_t=\dOz\Op_t$, leaving the master
relation
\begin{equation}
  R(t)=\frac{\sigma^2[U_t]}{\sigma^2[\hat A_t]}
  \;\simeq\;\frac{\Gamma(0)}{\ev{\rho_T^2}},
  \qquad
  \Gamma(\tau):=\ev{\dOz[Z_0]\,\dOz[Z_\tau]},
  \qquad \Gamma(0)=\Var[\Op_0],
  \label{eq:master}
\end{equation}
whose measured counterpart is shown in Fig.~\ref{fig:money} (right). The numerator is fixed by the baseline source variance; everything the method controls
sits in the denominator, which is a ledger of four contributions~\cite{ref:paper},
\begin{equation}
  \ev{\rho_T^2}
  =\underbrace{\Gamma(2T)}_{\text{horizon},\ \sim e^{-2\lambda_\star T}}
  +\underbrace{c(\dt)}_{\text{discretisation}}
  +\underbrace{\frac{\varsigma_f^2}{N_w}}_{\text{walkers}}
  +\underbrace{\frac{\sigma_\xi^2}{N_wN_h}}_{\text{trace probes}} .
  \label{eq:ledger}
\end{equation}
The first term is the autocorrelation of the source at twice the horizon; its
rate $\lambda_\star$ is the relaxation rate of the dynamics, a physical scale
with $\lambda_\star\sim\xi^{-2}$, so the horizon needed to reach a target
residual and the Monte Carlo time needed to decorrelate the chain grow together
towards the continuum limit; Fig.~\ref{fig:residual} (right) shows this decay
across a range of couplings in the free-theory case. The second is the $\dt$-plateau, visible in
Fig.~\ref{fig:money} (left) for the interacting theory. The third switches on only in the interacting theory, through the fluctuations of the
Hessian along the trajectory, and the fourth only if the trace in $\Stein f$ is
estimated with $N_h$ Hutchinson probes rather than exactly.\footnote{In Fig.~\ref{fig:residual}, the small volume considered allows an exact determination of the Jacobian trace, corresponding to the limit $N_h\to \infty$ in Eq.~\eqref{eq:ledger}.}

Once the horizon term has decayed below the floors the reduction saturates at a
value fixed by $c(\dt)$ alone. The resulting relation between the saturated
reduction factor and the measured plateau is a parameter-free consistency check
on the whole construction, which we develop in Ref.~\cite{ref:paper} together
with the ensembles, the tuning and the error analysis. What matters here is the
last term of Eq.~\eqref{eq:ledger}: evaluating $\Stein f$ exactly requires the
pointwise divergence, a $V$-fold forward-mode differentiation affordable for
benchmarking only, while Hutchinson probes reduce the cost to $\Op(V)$ at the
price of a floor that no increase of $T$ can remove. We return to this in
Sec.~\ref{sec:ml}.

\section{Route B: direct minimisation via effective neural propagator}
\label{sec:ml}
The second route parametrises $f_\theta$ and uses $\mathcal{F}$ of
Eq.~\eqref{eq:functional} itself, estimated on Monte Carlo samples, as the loss and gradient-based optimisation is used to find the optimal parameter $\theta$. This route was developed in detail in Ref.~\cite{ref:Spatscheck}, from
which we summarise the essential findings.

\begin{figure}[t]
  \centering
  \includegraphics[scale=0.25]{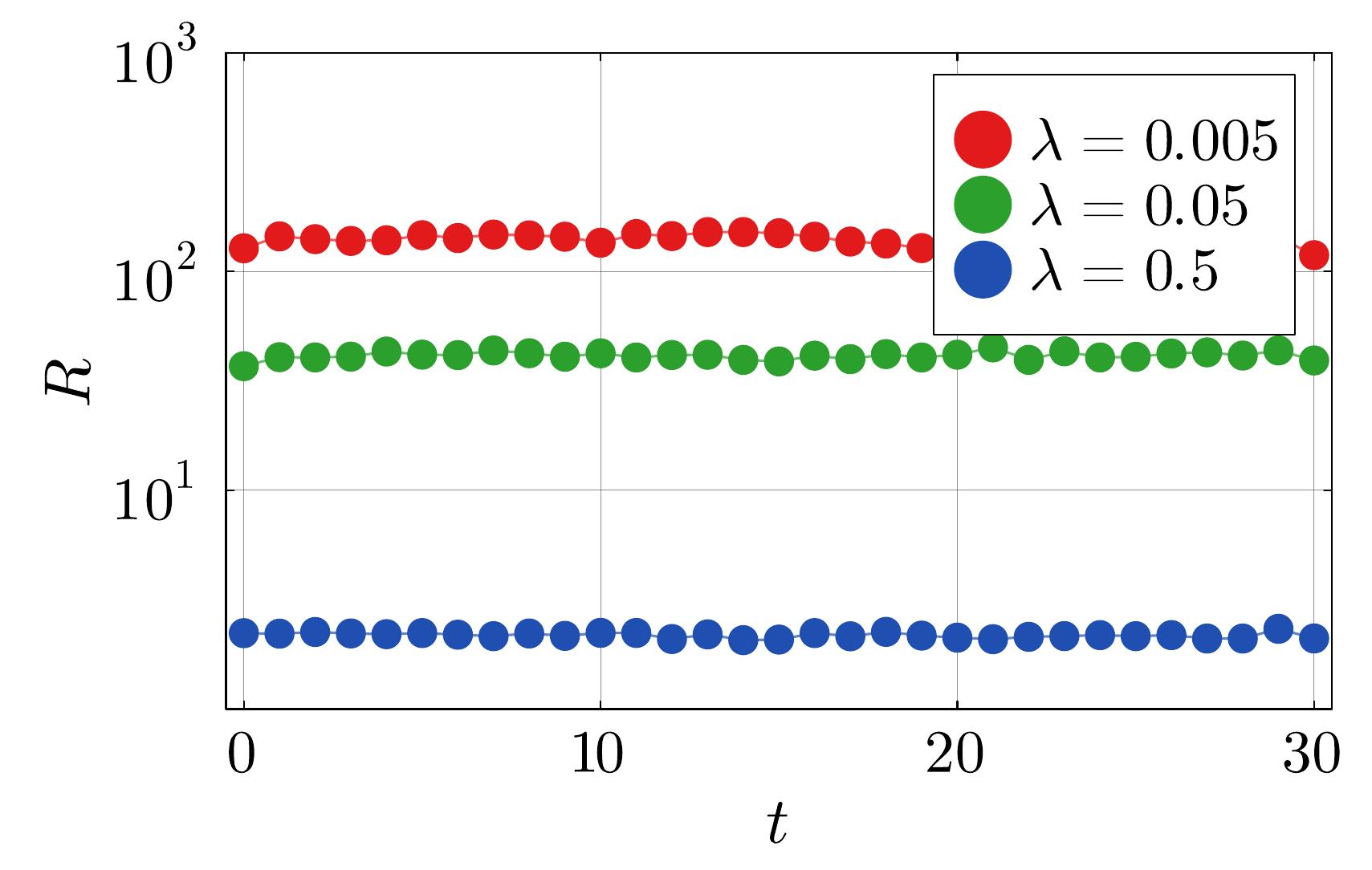}
  \includegraphics[scale=0.43]{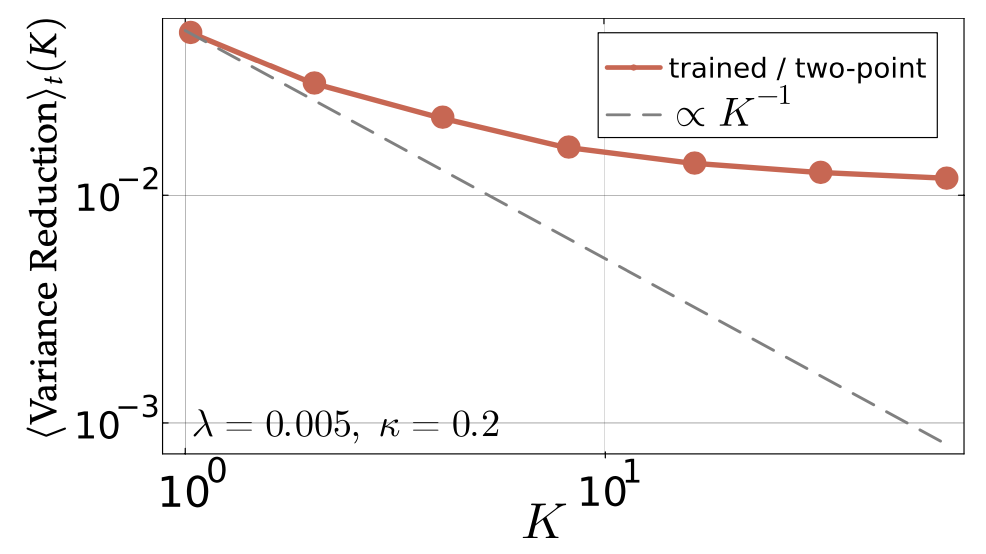}
  \caption{Direct minimisation, from Ref.~\cite{ref:Spatscheck}. \emph{Left:}
    Variance reduction factor for the $\lambda\phi^4$ theory at $\kappa=0.2$ for different value of $\lambda$
    \emph{Right:} reciprocal of time-averaged variance reduction against the number of
    Hutchinson probes. The dashed line is the $K^{-1}$ scaling expected when
    probe noise dominates; the departure marks the ceiling set by the quality of
    the learned $f_\theta$.}
  \label{fig:ml}
\end{figure}

The architecture is guided by the free-theory solution. Taking inspiration from
Fourier neural operators, $f_\theta$ is built from a momentum branch in which the networks $\mathbb{M}_\theta$ and $\mathbb{C}_\theta$, output a field-dependent self-energy dressing the free propagator, plus
a position-space convolutional branch,
\begin{equation}
  \tilde z = \mathrm{FFT}[z],\qquad f_\theta(z)=\mathrm{iFFT}\!\left[
  \frac{\delta(\mathbf{p})}{\hat p^2+\mathbb{M}_\theta(\tilde z)}\right]
  +\mathbb{C}_\theta(z),
  \label{eq:architecture}
\end{equation}
so that the exact solution is recovered when $\mathbb{M}_\theta$ is constant and $\mathbb{C}_\theta$ vanishes. In the free theory training converges cleanly with no overfitting and the weight distribution collapses from broad to sharply peaked, while the variance of the correlator falls by six to eight orders of magnitude, with the largest gains \emph{closest to criticality}, precisely where the standard estimator degrades most. In the interacting theory the gains shrink and are governed principally by the quartic coupling: a factor $\sim10^{2}$ at weak coupling, falling to marginal at the strongest coupling explored. Left panel of Fig.~\ref{fig:ml} shows such variance reduction factor for different ensembles.

The limitation is the trace. The divergence entering $\Stein f_\theta$ was
estimated with Hutchinson probes, whose relative error falls only as $K^{-1/2}$
while the cost grows linearly in $K$. Fig.~\ref{fig:ml} (right) shows the
consequence: the reduction tracks the expected $K^{-1}$ scaling at small probe
counts and then saturates, because past a certain accuracy the residual is set by
the quality of $f_\theta$ itself -- which is in turn limited by the same
stochastic noise entering the training loss. The bottleneck is not the
expressivity of the network but the estimator of a trace.

\section{Conclusions and outlook}
Source reweighting reformulates the StN degradation into an overlap problem, and the transport that closes it solves one pointwise equation, Eq.~\eqref{eq:steinpoisson}, whose operator generates Langevin dynamics. The stochastic resolvent and the trained flow are two ways of solving it; the estimator is unbiased for any transport, and the residual sets the gain through Eq.~\eqref{eq:master}.

Both routes meet the same obstruction, the trace in $\Stein f$. Route A needs it for $\rho_T$, and its stochastic estimate adds the last term of Eq.~\eqref{eq:ledger}; Route B needs it inside the loss, where probe noise caps the gain. The fix, in progress~\cite{ref:paper}, is architectural: coupling-layer flows have triangular Jacobians and closed-form log-determinants, so exact divergence becomes a property of the parametrisation. On the other hand, Route A can be improved via higher-order PDE integration schemes, and by better leveraging Markov semigroup theory

The final target, however, is gauge theory, where the flow becomes the exponential map of Sec.~\ref{sec:flow-exp}. 
Not every construction survives there: those that invert the Hessian globally inherit its spectrum, which has exact zero modes from gauge redundancy and, by compactness, cannot be positive everywhere. Nevertheless, a stochastic approach like route A (or a different decomposition of the Stein operator) could extend this method to gauge fields. Likewise, adapting existing gauge-equivariant NF architectures to route B could provide a way to attempt direct KL minimsation.

\acknowledgments
The work of P.~B. is supported by the Carlsberg Foundation, grant CF22-0922.
A.~N. acknowledges support by the Simons Foundation grant 994300 (Simons Collaboration on Confinement and QCD Strings) and from the SFT Scientific Initiative of INFN. We thank CINECA for access to the LEONARDO supercomputer under the CINECA-INFN agreement.


\end{document}